\documentclass[aps,groupedaddress,showpacs,twocolumn]{revtex4}
\usepackage[dvips]{graphicx}            
\usepackage[]{caption}
\usepackage{amsmath}
\usepackage{amssymb}
\newcommand\lsim{\mathrel{\rlap{\lower4pt\hbox{\hskip1pt$\sim$}}
    \raise1pt\hbox{$<$}}}
\newcommand\gsim{\mathrel{\rlap{\lower4pt\hbox{\hskip1pt$\sim$}}
    \raise1pt\hbox{$>$}}}

\begin{document}

\bibliographystyle{prsty}
\title{Large--scale magnetic fields from density perturbations} 
\author{S. Matarrese$^{1,2}$, S. Mollerach$^3$, A. Notari$^{4,5}$ and A. 
Riotto$^2$}
\affiliation{(1) Dipartimento di Fisica `G. Galilei', Universit\`a di 
Padova, via Marzolo 8, I-35131 Padova, Italy}
\affiliation{(2) INFN, Sezione di Padova, via Marzolo 8, I-35131 Padova, Italy}
\affiliation{(3) CONICET, Centro At\'omico Bariloche, 
Av. Bustillo 9500, Bariloche, Argentina}
\affiliation{(4) Scuola Normale Superiore, Piazza dei Cavalieri 7, 
I-56126 Pisa and INFN, Pisa, Italy}
\affiliation{(5) Physics Department, McGill University, 3600 University Road,
Montr\'eal, QC, H3A 2T8, Canada} 
\date{\today}

\begin{abstract}
\noindent
We derive the {\it minimal} seed magnetic field which unavoidably arises in 
the radiation and matter eras, prior to recombination, 
by the rotational velocity of ions and electrons,
gravitationally induced by the non-linear evolution of primordial 
density perturbations. 
The resulting magnetic field power-spectrum is fully determined by 
the amplitude and spectral index of density perturbations. 
The {\it rms} amplitude of the seed-field at recombination is 
$B \approx 10^{-23} (\lambda/{\rm Mpc})^{-2} $ G, on comoving scales 
$\lambda \gsim 1$ Mpc.  
\end{abstract}

\pacs{98.80.cq, preprint DFPD 04/A-25}

\maketitle

\section{Introduction}

Magnetic fields are present in most astrophysical systems 
\cite{obsreview,Torn,Grasso,Widrow,Carilli}, 
but their origin is still unknown. 
Spiral galaxies typically contain magnetic fields of about $10^{-6}$ G 
that are aligned with the spiral density waves \cite{obsreview}. 
A plausible explanation is that these fields have been produced from the 
exponential amplification of an initially weak
field by a mean-field dynamo \cite{dynamo,moffatt}, 
in which a seed-field was amplified by the differential
rotation of the galaxy in conjunction with magnetohydrodynamic turbulence.
Magnetic fields of comparable amplitude are also found on cluster scales 
\cite{bolo}. 

The seed-field strength required at the time of completed galaxy formation
for a galactic dynamo to produce the present magnetic field strength
of a few $\mu$G is usually quoted in the range
$\sim (10^{-23}$ -- \mbox{$10^{-19}$) G} 
\cite{obsreview,Torn,Grasso,Widrow,Carilli}. 
These estimates, though, are obtained by considering the dynamo 
amplification in a flat Universe with zero cosmological constant for
``typical'' values of the parameters of the $\alpha\omega$--dynamo. 
According to Ref.~\cite{dlt} these lower bounds can be relaxed 
to about $10^{-30}$ G, for a
Universe with a dark energy component (e.g. a cosmological constant or
quintessence), which appears to be favored by recent results from
high-redshift Supernovae observations, cosmic microwave background  
experiments and large--scale structure  data. 


Most proposed models of primordial seed-field generation either fail to meet 
these requirements or invoke non-standard coupling between the electromagnetic 
field and the inflaton scalar sector responsible for a primordial 
period of accelerated Universe expansion (see, e.g. Ref.~\cite{Widrow} 
and references therein). 

The most economic and conservative physical mechanism for producing such a 
seed-field was proposed by Harrison
\cite{Harrison}. The mechanism relies on the fact that weak magnetic fields 
are generated during the radiation era in regions that have non-vanishing 
vorticity.
The main problem with Harrison's mechanism is that the required vorticity 
of the plasma does not have any dynamical origin, but has to be put in 
by hand, as an initial condition. 
In this paper we show that rotational velocity of the plasma 
is unavoidably induced gravitationally by the nonlinear mode-mode coupling 
of primordial density perturbations generated during inflation, arising 
at second order in perturbation theory.  
Therefore, we predict that a minimal seed magnetic field is generated  by 
the differential rotational velocity of ions and electrons. 
The resulting magnetic field power-spectrum is fully determined by the  
amplitude and the spectral index $n_S$ of primordial scalar 
perturbations. The {\it rms} amplitude of the seed-field at 
recombination is $B \approx 10^{-23} (\lambda/{\rm Mpc})^{-2}$ G,
on comoving scales $\lambda \gsim$ Mpc. 

The plan of the paper is as follows. 
In Section II we briefly recall Harrison's mechanism and we show how the 
seed magnetic field is produced. In Section III we compute the power-spectrum 
of such a seed-field in terms of the power-spectrum of primordial scalar 
perturbations generated during inflation. Finally, in Section IV we draw our 
conclusions. 

\section{Generation of seed magnetic field} 

Let us first recall how Harrison's
mechanism \cite{Harrison} works. Consider a rotating region in the expanding
early Universe (with scale factor $a$), consisting of matter (mostly
protons) with average energy density $\rho_m$, electrons and photons
with energy density $\rho_{\gamma}$. Electrons and photons are
tightly coupled and are considered as a single fluid in Ref.~\cite{Harrison}.
Let $\omega_m$ and  $\omega_{\gamma}$ be their angular velocities.
In absence of interactions, their angular momenta, proportional 
respectively to $\rho_m \omega_m a^5$, and $\rho_{\gamma} \omega_{\gamma} 
a^5$, are separately conserved (see also Appendix \ref{Kelvin}).  
As the Universe expands, they scale differently,
$\omega_m \propto  a^{-2}$ and $\omega_{\gamma} \propto a^{-1}$.  
Ions and the electron-photon fluid spin down at different
rates and then currents and magnetic fields are generated.

A formal derivation of the resulting magnetic field strength in terms
of the vorticity in the proton fluid including the electron-proton
coupling can be found in \cite{Widrow}. It is necessary to
consider the evolution of the multicomponent fluid formed by photons,
electrons and protons during the radiation era.
At temperatures $T\gtrsim m_e$ (where $m_e$ is the electron mass), the
interactions between ions and electrons are strong due to copious 
lepton-pair production, and they are locked
together. The interaction between electrons and photons
is also very strong. This means that all the plasma has the same angular 
velocity at $T \approx m_e$ and no magnetic field can be generated.  
Below this temperature, electrons and photons
are still tightly coupled through Thomson scattering. Protons and electrons
are tightly coupled through Coulomb scattering 
(scattering between
photons and protons can be neglected as the coupling is weaker) and so the 
photon fluid drags the protons in its motion. 
The difference in the strength of these interactions and in the mass
of electrons and protons leads however to a
small difference in the electron and proton fluid rotational 
velocities, that gives rise to non-vanishing currents and magnetic fields. 
In our case the rotational velocity of the plasma is sourced by the 
non-linear evolution of scalar (density) perturbations.  
The generation of the magnetic field 
ends at recombination, when electrons and protons
combine to form neutral hydrogen and radiation decouples from matter.
This means that the generation of the magnetic field 
starts around $T\approx m_e$ and ends when 
$T \approx T_{rec}$. 
Dark energy and neutrinos can be ignored throughout 
the evolution. Moreover, during the radiation era
we can safely neglect the role of the Cold Dark Matter (CDM) 
component, that is, we consider a plasma with only baryons 
(protons), electrons and photons, and electromagnetic (EM) fields.
After matter-radiation equality, CDM plays the dominant 
gravitational role and it enters our equations determining the 
evolution of perturbations. 

Let us now go into the details of this mechanism. 
The momentum equation for the interacting components can be written using
the total energy-momentum conservation equation $T^\nu_{\mu;\nu}=0$
The energy-momentum tensor of each component is not conserved 
independently and its divergence has a source term that takes into
account the energy and momentum transfer among the components,
$T^{(A)\nu}_{\mu;\nu}=Q^{(A)}_{\mu}$, with $A=\gamma,
p, e, EM$. We will describe the proton, electron and photon components
as approximately perfect fluids, thus their 
energy momentum tensor can be written as
\begin{equation}
T^{(\alpha)\mu}_{\nu}=(p^{(\alpha)}+\rho^{(\alpha)}) u^{(\alpha)\mu} 
u_{(\alpha)\nu}+ p^{(\alpha)}
\delta^{\mu}_{\nu}  \hspace{0.4cm} ,  
\end{equation}
with $\alpha=e,p,\gamma$ and $p^{(\alpha)}=w^{(\alpha)}\rho^{(\alpha)}$. 
We can expand the energy density and four velocity of each component as:
\begin{eqnarray}
\rho^{(\alpha)}&=&\rho^{(\alpha)}_0 \left( 1+\delta^{(\alpha)}
\right) \;,
\nonumber\\
u^{(\alpha)\mu}&=&\frac{1}{a}
\left(
\delta^{\mu}_0+ v^{(\alpha)\, \mu}\right) \;.
\end{eqnarray}
In what follows we will neglect the electron and proton pressure, i.e.
we set $w^{(e)} \approx w^{(p)} \approx 0$, while for the photons 
we have $w^{(\gamma)} =1/3$. 

The crucial step in our computation is that 
we have to include second-order 
terms in the metric and the matter perturbations 
\cite{mmb,mhm,reviewNG}, since rotational component of the velocity, 
and hence magnetic field, are only generated at second order. 
Indeed linear vector modes are not generated during inflation and,
by Kelvin's circulation theorem vorticity is conserved for 
a perfect fluid. Of course, this conservation holds at second order as well, 
and it applies to the vorticity of the fluid which generates the gravitational 
field. As explained in detail in Appendix \ref{Kelvin}, this 
conservation law does not prevent the occurrence of a rotational  
velocity component of the charged particles, thus giving rise to a 
non-zero magnetic field. 

Let us expand 
$\delta^{(\alpha)}=\delta^{(\alpha)}_{(1)}+\delta^{(\alpha)}_{(2)}$ and 
$v^{(\alpha)\, \mu}=v^{(\alpha)\, \mu}_{(1)}
+v^{(\alpha)\, \mu}_{(2)}$, where the superscripts $(1)$ and $(2)$
indicate first and second-order perturbation quantities. We will
work in the Poisson gauge, which is the generalization 
of the so-called longitudinal (or Newtonian) gauge when vector and
tensor modes are allowed. The perturbed
metric reads 
\begin{eqnarray}
 ds^2 &=& a^2(\eta)\left\{-(1+2\phi)d\eta^2
+ 2 \chi_i  d\eta dx^i +\right. \nonumber\\
 &+& \left.\left[ 
(1-2\psi) \delta_{i j}+ \chi_{i j} 
\right]  dx^i dx^j \right\} .
\end{eqnarray}
In this gauge, the first-order scalar perturbations are equal if there
are no anisotropic stresses,
$\phi^{(1)}=\psi^{(1)}=\varphi$.
The vector perturbation is $\chi_i$, and we will
assume that it is not generated at first order
$\chi_i^{(1)}=0$. We will also neglect primordial tensor modes.

Although first order primordial vector and tensor modes are absent,
the non-linear evolution of the primordial scalar perturbations gives 
rise to non-vanishing vector and tensor perturbations (and corrections
to the scalar perturbations) at second order,
$\chi_i=\chi_i^{(2)}$ and $\chi_{ij}=\chi_{ij}^{(2)} 
$\cite{mmb,mhm}. 

Let us consider the momentum continuity equation for each
fluid component (whose general expression up to second order in 
perturbation theory is reported in Appendix \ref{Kelvin}). 
The contribution of the
electromagnetic field can be included in the source of the
charged fluid components, as $T^{(EM)\nu}_{\mu;\nu}=
F_{\mu\beta}j^{\beta}$, with $j^{\beta}\equiv e n (u^{(p)\beta}-
u^{(e)\beta})$, where we have assumed charge neutrality
 ($n=n^{(e)} \simeq n^{(p)}$). 
The momentum equation for photons up to second order 
can be written as 
\begin{eqnarray}
&&\frac{4\rho_0^{(\gamma)}}{3} 
\left((v^{(\gamma)}_{i}+\chi_i)'+\frac{1}{4} 
\partial_i\delta^{(\gamma)}+\partial_i\phi+\frac{1}{4}(3
\varphi 
-\delta^{(\gamma)})\partial_i\delta^{(\gamma)}
\right. \nonumber \\
&&- \varphi'v^{(\gamma)}_{i}-\left.\frac{1}{3}  v^{(\gamma) ,j}_{j}
v^{(\gamma)}_{i}
 + \frac{1}{2}(v^{(\gamma)2})_{,i}+ \frac{1}{2}(\varphi^2)_{,i}
\right)=\kappa_i^{e\gamma} \;,
\label{photoneq}   
\end{eqnarray}
and for protons
\begin{eqnarray}
&\rho_0^{(p)}&\left((v^{(p)}_{i}+\chi_i)' + {\cal H}(v^{(p)}_{i}+\chi_i)
+\partial_i\phi \right.\nonumber \\
&-&\left.2\varphi'v^{(p)}_{i} 
+\frac{1}{2} (v^{(p)2})_{,i}+\frac{1}{2} (\varphi^2)_{,i}
\right) \nonumber \\
&=& e n(E_i+\epsilon_{ijk}v^{(p)i}B^k)-\kappa_i^{ep} \;,
\label{protoneq}
\end{eqnarray}
where ${\cal H}=a^\prime/a$ is the expansion rate in conformal time.  
The momentum equation for electrons is similar to that for
protons, but it has the opposite sign in the righ-hand side terms and an
additional source, $-\kappa_i^{e\gamma}$, due to the momentum transfer
between the electron and photon fluids. This momentum transfer is
due to Thomson scattering and is given by $\kappa_i^{e\gamma}=
-\frac{4}{3}\rho_0^{(\gamma)}\tau'(v^{(\gamma)}_{i}-v^{(e)}_{i})$, where
the differential optical depth is $\tau'=a n \sigma_T$, and $\sigma_T$
is the Thomson cross section. The momentum transfer between electrons
and protons is due to Coulomb scattering and  can be written as 
$\kappa_i^{ep}=-n (v^{(p)}_{i}-v^{(e)}_{i})/\tau_e$, where the
collision time between electrons and protons is $\tau_e=m_e\sigma/ne^2$ in
terms of the conductivity of the plasma $\sigma$.

During the whole period we are interested in, up to the time
of hydrogen recombination, the collision time between electrons and
photons ($\propto \tau'^{-1}$) and the electron-proton one are
much shorter than the Universe expansion time, thus momentum transfer
is very efficient. A tight coupling expansion of the momentum
equations gives, to lowest order, $v^{(\gamma)}_{i} \simeq v^{(e)}_{i} \simeq
v^{(p)}_{i}$. 

We can obtain an equation for the vorticity of the proton fluid by
taking the curl of Eq.~(\ref{protoneq}) and combining it with 
Maxwell's equations
\begin{equation}
\frac{d}{d\eta} \left[a^2(\zeta^{(p)}_{i}+\Omega_i
+\frac{e}{m_p}B_i)\right]=2 \epsilon_{ijk} a \varphi'^{,j}v^{(p)k}
+\frac{e a^{2}}{\sigma m_p}\nabla^2B_i,
\label{curlproton}
\end{equation}
where we have kept terms up to second order in the metric perturbation
and defined $\zeta^{(p)}_{i}\equiv\epsilon_{ijk}\partial^j v^{(p)k}/a$ and 
$\Omega_i\equiv\epsilon_{ijk}\partial^j \chi^k/a$. 
In the last equation the diffusion term can be dropped in the highly
conductive protogalactic medium.

The seed magnetic field can be written as
\begin{equation}
B_i=-\frac{m_p}{e}\left(\beta^{(p)}_{i}-\frac{a_I^2}{a^2}
\beta^{(p)}_{(I)\,i} 
-\frac{2}{a^2}\int^\eta_{\eta_I} d\tilde{\eta} a 
\epsilon_{ijk}\partial^j\varphi'v^{(p)k}\right) \;,
\label{risultB}
\end{equation}
where the subscript ``$I$'' denotes the initial time, corresponding to 
$T\sim m_e$, when $B_I=0$, and where we 
introduced the useful combination 
$\beta^{(A)}_i\equiv \zeta^{(A)}_i+\Omega_i$. 

Taking now the curl of the photon momentum equation (\ref{photoneq}) 
we obtain an equation for the photon vorticity
\begin{eqnarray}
(a \beta^{(\gamma)}_i)' &=&
\frac{3}{4}\epsilon_{ijk}
\left[-\partial^j\varphi \partial^k\delta^{(\gamma)}
 \right. \nonumber \\
&-& \left.
 \frac{4}{9}v^j
\partial^k\partial^lv_l+\frac{4}{3}\partial^j\varphi'v^k\right],
\label{curlp}
\end{eqnarray}
where we have set $v^{(\gamma)} \equiv v$. 
Solving Eq.~(\ref{curlp}) we finally get:
\begin{eqnarray}
\beta_i(\eta) &=& \frac{a_I}{a}\beta_{(I)\,i}+ \frac{3}{4 a} 
\int_{\eta_I}^{\eta}  d\tilde{\eta}
\epsilon_{ijk} \left(
-\partial^j\varphi \partial^k\delta^{(\gamma)}  \right. \nonumber \\
&-& \left. \frac{4}{9}v^j 
\partial^k\partial^l v_l + \frac{4}{3}
\partial^j \varphi' v^{k} \right),  \label{beta}
\end{eqnarray}
where we have set $\beta^{(p)}\simeq \beta^{(\gamma)} \equiv \beta$.  

Equation (\ref{risultB}) with the solution Eq.~(\ref{beta}) is our main 
result and provides the resulting magnetic field in terms of the metric 
and matter perturbations. In order to finalize our computation 
we have to consider the linearly perturbed Einstein's equations. For a
Universe dominated by a fluid with equation of state
$p= w \rho$, with $w={\rm const.}$, we can express the fluid velocity 
$v^{(1) \, i}$ and the
energy density perturbations $\delta^{(1)}$ in terms of the
gravitational potential $\varphi$ as
\begin{eqnarray}
v^{(1) \, i}&=&-\frac{2}{3(1+w) {\cal H}^2  } 
\partial _i(\varphi'+{\cal H} \varphi) \;,\nonumber
\\
\delta^{(1)}&=&\frac{2}{3 {\cal H}^2}[\nabla^2\varphi-3 
{\cal H}(\varphi'+{\cal H}\varphi)]  \, .   \label{primoordine}
\end{eqnarray}
The evolution equation for the peculiar gravitational potential
$\varphi$ is given by 
\begin{equation}
\varphi''+ 3{\cal H}(1+w)\varphi'-w \nabla^2\varphi=0 \;,
\label{eqmoto}
\end{equation}
whose solution in the radiation dominated era in Fourier space reads 
\begin{equation}
\label{stagn}
\varphi({\bf k},\eta) = \frac{3 j_1(x)}{x} 
~\varphi_0 ({\bf k}) \;, 
\end{equation}
where $x\equiv k \eta/\sqrt{3}$ and $j_\ell$ denote spherical Bessel 
functions of order $\ell$. 
The latter expression manifests the well-known stagnation effect 
for perturbations which crossed the Hubble radius during the radiation
era.  

We can now evaluate in Eq.~(\ref{risultB}) the contribution from 
$\beta(\eta)$, given by Eq.~(\ref{beta}). With 
the solutions Eqs.~(\ref{primoordine}) it reads 
\begin{eqnarray}
\beta_i(\eta) & = & \frac{a_I}{a}\beta_{(I)\,i} + 
\frac{1}{a}\int_{\eta_I}^{\eta} d\tilde{\eta} \epsilon_{ijk} \left(
\frac{2 \partial^j \varphi \partial^k \varphi'}{ {\cal H}}
\right. \nonumber \\
&-& \left. \frac{ 7 \partial^j \varphi \nabla^2 \partial^k 
\varphi }{12 {\cal H}^2}-
\frac{  \partial^j \varphi \partial^k \nabla^2 \varphi' }{12 {\cal H}^3} 
\right. \nonumber \\
&-& \left. 
\frac{ \partial^j \varphi' \partial^k \nabla^2 \varphi }{12 {\cal H}^3}-
\frac{ \partial^j \varphi' \partial^k \nabla^2 \varphi' }{12 {\cal H}^4}
\right)    \label{betaespl}
\end{eqnarray}
This expression can be integrated analytically exploiting 
the non-trivial result given by Kelvin's circulation theorem 
(see Appendix \ref{Kelvin}), which states the conservation of angular 
momentum to all orders. In fact, it can be seen that at second order
\begin{eqnarray}
\beta_i(\eta)&=& \frac{1}{a} \epsilon_{ijk}\left( 
\frac{2}{{\cal H}^2} \partial^j \varphi'\partial^k \varphi
-\frac{1}{12 {\cal H}^3} \partial^j \varphi \partial^k \nabla^2 \varphi 
\right. \nonumber\\
&-&\left. \frac{1}{12 {\cal H}^4} \partial^j \varphi' 
\partial^k \nabla^2 \varphi
 \right)
\label{betaintegrato}
\end{eqnarray}
where we simply used Eqs.~(\ref{primoordine},\ref{eqmoto}), with $w=1/3$, 
in Eq.~(\ref{omega2}), assuming vanishing vorticity of the radiation
fluid, $\omega_i=0$, which fixes the value of our integration constant
$\beta_{(I)\,i}$. 
One can explicitly check that the time derivative of 
Eq.~(\ref{betaintegrato}) gives the integrand in Eq.~(\ref{betaespl}).

\section{Power-spectrum of the seed magnetic field}

We can now compute the power-spectrum of the magnetic field generated
up to recombination. The correlation of the Fourier modes
of the field can generally be written as \cite{du03} 
\begin{eqnarray}
\langle B_l({\bf k})B_m^\ast({\bf k'})\rangle&=&\frac{(2\pi)^3}{2}
\delta^3({\bf k}- {\bf k'})[(\delta_{lm}-\hat k_l\hat k_m)S(k)
\nonumber\\   
&+& i \epsilon_{lmj} \hat{k}_j A(k)]~.  \label{ansatz}
\end{eqnarray}
where $\hat {\bf k}$ denotes the unit vector in direction of ${\bf k}$. 
The term proportional to $A(k)$ represents a ``helical'' component,
that is a non-vanishing field component in the direction of the
current (${\bf B}\cdot({\bf \nabla}\times{\bf B} \neq 0$)). It can be
seen that the seed field obtained in the previous section has no
helical component, thus $A(k)=0$. The relevant component $S(k)$ can
be obtained from
\begin{equation}
\langle {\bf B}({\bf k})\cdot {\bf B}^\ast({\bf k'})\rangle=(2\pi)^3
S(k)\delta^3({\bf k}- {\bf k'}).
\end{equation}
By replacing Eq.~(\ref{betaintegrato}) into Eq.~(\ref{risultB}), the
Fourier modes of ${\bf B}$ can be written as 
\begin{eqnarray}
&&{\bf B}({\bf k},\eta)=-\frac{m_p (1+z)}{e {\cal H}^2} 
\int \frac{d^3 k'}{(2\pi)^3} {\bf k}\times{\bf k'}\nonumber\\
&&\left[2 \varphi'(|{\bf k}-{\bf k'}|)\varphi(k')
-\frac{k'^2}{12{\cal H}^2}\varphi'(|{\bf k}-{\bf k'}|)\varphi(k')
\right.\nonumber\\
&&\left.-\frac{k'^2}{12{\cal H}}\varphi(|{\bf k}-{\bf k'}|)\varphi(k')
\right],\label{bk}
\end{eqnarray}
where we have neglected the last term in Eq.~(\ref{risultB}).

Up to the matter-radiation equality time the gravitational potential 
and its derivative are given by
Eq.~(\ref{stagn}), and the correlation function of ${\bf B}$ modes can be
computed in terms of that of $\varphi_0({\bf k})$,
which is a Gaussian random field with auto-correlation function
\begin{equation}
\langle\varphi_0({\bf k})\varphi_0({\bf k'})\rangle=(2\pi)^3 P_\varphi(k)
\delta^3({\bf k}- {\bf k'}),
\end{equation}
where $P_\varphi(k)$ is the gravitational potential 
power-spectrum, $P_\varphi(k)= P_{0\varphi} k^{-3} (k/k_0)^{n_s-1}$ 
and $k_0$ is some pivot wave-number.
For the normalization of $P_\varphi$ we can use its relation 
with the power-spectrum of the
comoving curvature perturbation (see e.g. Ref.~\cite{pe03}) 
$\cal R$, namely $\Delta^2_{\cal R}(k) = \Delta^2_{\cal R}(k_0) 
(k/k_0)^{n_s-1}$, where $\Delta^2_{\cal R}(k_0) = (25/9) 
(P_{0\varphi}/2\pi^2)$ 
\footnote{The relation with the inflation
parameters is $\Delta^2_{\cal R}=H^2/(\pi\epsilon m_P^2)$, with 
$m_P\equiv G^{-1/2}$}.
Then, the resulting magnetic field is a {\it chi-square} distributed
random field with power-spectrum 
\begin{eqnarray}
S(k,\eta)&=&\frac{162m_p^2(1+z)^2}{e^2{\cal H}}
\int\frac{d^3 k'}{(2\pi)^3} 
P_\varphi(|{\bf k}-{\bf k'}|)P_\varphi(k')\nonumber\\
&&|{\bf k}\times{\bf k'}|^2
\left[\frac{x_1}{4x_2}j_1(x_1)j_1(x_2)\right.\nonumber\\
&&\left.+j_2(x_2)j_1(x_1)\left(\frac{2}{x_1}-\frac{x_1}{4}\right)
\right]^2
,\label{sk}
\end{eqnarray}
where we have defined $x_1=k'\tilde{\eta}/{\sqrt 3}$, 
$x_2=|{\bf k}-{\bf k'}| \tilde{\eta}/{\sqrt 3}$.  

The main contribution to the power-spectrum comes from the term in
${\bf B}$ with a time derivative of $\varphi$ (second term in
Eq.~(\ref{sk})). As $\varphi$ tends to a constant 
after equality, the contribution to the generation of ${\bf B}$ during the
period from equality to recombination is subdominant with respect to
that up to equality time. 

\begin{figure}[t]
\includegraphics[width=9cm]{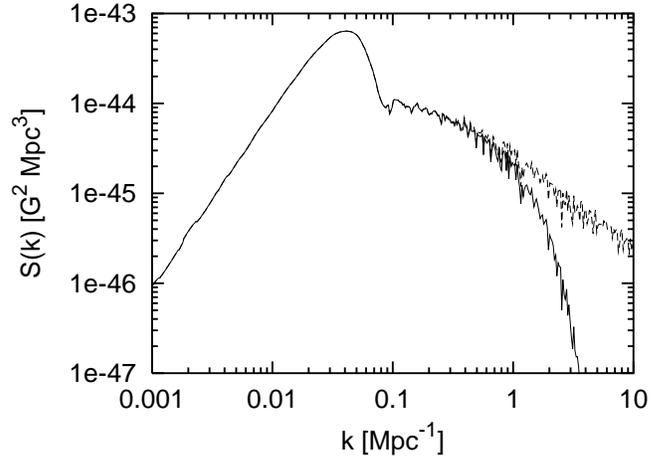}\\
\caption[fig0]{\label{fig1} 
Power-spectrum of the magnetic field at recombination as a function
of the comoving wavenumber. The upper and lower curves refer to the case 
with or without Silk-damping included, respectively.}
\end{figure}

We show in Figure 1 the power-spectrum of ${\bf B}$ at recombination, 
$S(k)$, obtained from the numerical integration of  Eq.~(\ref{sk}),
as a function of the comoving wavenumber $k$. We have used the  following 
values for the model parameters: $\Delta^2_{\cal R}(k_0)=2.3\times10^{-9}$,
$z_{\rm EQ}=3454$, $z_{\rm rec}=1088$, $h=0.7$ and $n_s=1$
\cite{sp03,pe03}. At small scales diffusion damps the fluctuations in
the photon and baryon fluids. This effect can be taken into account by
a factor $\exp (-k^2/k_D^2)$ multiplying the velocity perturbations (see
e.g. \cite{do03}). As shown in the plots, 
Silk damping affects the produced magnetic field 
on scales smaller than about 2 comoving Mpc. 

For small $k$ ($k<0.04$ Mpc$^{-1}$), $S(k)\propto k^2$, as pointed out
in Ref.~\cite{du03}, in order for the magnetic field to be divergenceless.
For wavenumbers $k>1$ Mpc$^{-1}$ the magnetic field spectral index approaches
instead $n\simeq -1$, i.e. $S(k)\propto k^{-1}$, for scale-invariant ($n_S=1$) 
primordial perturbations. For a general scalar spectral index $n_S$, 
$S(k)\propto k^{2n_S-3}$. 

The mean square value of the field on a given scale $\lambda$ is obtained by
averaging over a volume of size $\propto \lambda^3$ \cite{hi97,du03},
convolving with a Gaussian window function, namely 
\begin{equation}
B_\lambda^2=\int\frac{d^3k}{(2\pi)^3} S(k) \exp(-\lambda^2k^2/2).
\end{equation}
For comoving scales larger than $\sim$ 1 Mpc, up to 50 Mpc 
the {\it rms} magnetic field at recombination is well approximated by 
\begin{equation}
B_\lambda(\eta_{\rm rec}) \approx 10^{-23} (\lambda/{\rm Mpc})^{-2} 
\, {\rm G}.
\end{equation}
After recombination the magnetic field can be considered to be
frozen into the plasma and thus it redshifts with the expansion of the
Universe as ${\bf B}(\eta)={\bf B}(\eta_{\rm rec}) (a/a_{\rm rec})^{-2}$. Then,
the average field scaled to its value today results
\begin{equation}
B_\lambda(\eta_0) \approx 10^{-29} (\lambda/{\rm Mpc})^{-2} \, {\rm G}.
\end{equation}

\section{Conclusions}

In this paper we have discussed a new mechanism for the generation 
of the cosmic seed magnetic field. It 
acts during the evolution of the Universe 
up to the epoch of hydrogen recombination. 
The underlying physical phenomenon is the vorticity of charged particles 
driven by gravitational vector modes. The latter, in turn, arise from the 
non-linear evolution of purely scalar (density) primordial perturbations. 
Hence, our mechanism is a generic prediction of the standard 
hierarchical structure formation scenario and does not require any ad-hoc 
assumption.  

Maybe the most important feature of the created magnetic field discussed 
in this paper is that its power-spectrum is fully determined by 
the power-spectrum of the primordial density perturbations.
As a consequence, a significant signal is expected over 
cosmological scales much greater than those encountered in 
other mechanisms acting during the early evolution of the Universe. 


\begin{acknowledgments}

We would like to thank Klaus Dolag, Dario Grasso, Diego Harari
 and Igor Tkatchev for discussions. 

\end{acknowledgments}

\appendix  
\section{Vorticity conservation} \label{Kelvin}

We want to illustrate here how one can generate a non-zero rotational 
component of the fluid velocity at second order in the (dominant) 
radiation fluid without violating the conservation of vorticity, 
dictated by Kelvin's circulation theorem, which is non-perturbative and 
applies to a perfect fluid with equation of state $p=p(\rho)$ coupled to 
gravity (see Ref.~\cite{Ellis}).

As we can see from Eq.~(\ref{curlp}), we were indeed able to generate at 
second order a non-zero value for $\beta_i^{(\gamma)}$ (and also for  
$\zeta_i^{(\gamma)}$). And then, through Eq.~(\ref{risultB}), this 
$\zeta_i\simeq\zeta_i^{(p)}\simeq\zeta_i^{(\gamma)}$ acts as a source for 
the magnetic field. The crucial point is that none of these quantities 
(neither $\zeta_i$ nor $\beta_i$) coincides at second order with the quantity 
that is conserved, according to Kelvin's theorem.  

We recall what the theorem states \cite{Ellis}. First of all one considers a 
system with a single perfect fluid (in our case this is the radiation 
fluid, in the limit in which the other components are negligible), 
with four-velocity $u^{\mu}$.
Then one constructs the quantity
\begin{equation}
v_{\mu\nu}=h_{\mu}^{\, \alpha} h_{\nu}^{\, \beta} u_{\alpha ;\beta} \, ,
\end{equation}
where $h_{\mu \nu}\equiv g_{\mu \nu}+u_{\mu} u_{\nu}$ is the projection 
tensor into the rest-frame of an observer moving with 4-velocity $u^{\mu}$.
Then we split $v_{\mu \nu}$ it into its symmetric and antisymmetric parts,
namely
\begin{equation}
v_{\mu \nu}=\theta_{\mu \nu} + \omega_{\mu \nu} \, ,
\end{equation}
where the antisymmetric part $\omega_{\mu \nu}$ is called the {\it 
vorticity tensor}.
The symmetric part $\theta_{\mu \nu}$ can be further split into its trace 
({\it volume expansion}) and trace-free part ({\it shear tensor}):
\begin{equation}
\theta_{\mu\nu}\equiv\sigma_{\mu\nu}+\frac{1}{3} \theta h_{\mu \nu}  \, 
\hspace{6mm}  \theta= u^{\mu}_{\phantom{\mu} ;\mu} \, .
\end{equation}
Finally, one can construct the {\it vorticity vector} as 
\footnote{Here $ \eta^{\mu \nu \rho \sigma}$ is the totally 
antisymmetric Levi-Civita tensor, 
with $\eta^{1234}\equiv (-g)^{-\frac{1}{2}}$ }:
\begin{equation}
\omega^{\mu}\equiv \frac{1}{2} \eta^{\mu \nu \rho \sigma} u_{\nu} 
\omega_{\rho \sigma} \, .
\end{equation}

It is possible to show now, using the field equations \cite{Ellis}, 
that this quantity obeys the equation 
\begin{equation}
h^{\mu}_{\phantom{a} \nu}
(l^2 \omega^{\nu}\dot{)}=\sigma^{\mu}_{\phantom{a} \nu} 
(l^2 \omega^{\nu})+ \frac{l^2}{2} 
\eta^{\mu \nu \rho \sigma}u_{\nu}\dot{u}_{\rho ;\sigma} \, ,  \label{eqomega}
\end{equation}
where the overdot stands for the convective time derivative along
$u^\mu$ and the scalar $l$ is defined by
\begin{equation}
\frac{\dot{l}}{l}=\frac{1}{3} \theta  \, ,  
\end{equation} 
thus representing the local expansion of the fluid in its rest frame
(which coincides 
with the scale-factor in an exactly homogeneous and isotropic FRW model).
Substituting into Eq.~(\ref{eqomega}) the momentum-conservation equation 
one can find (see Ref.~\cite{Hwang}) that the 
following equation holds for $\omega\equiv\sqrt{\omega_{\mu}\omega^{\mu}}$:
\begin{equation}
\frac{\dot{\omega}}{\omega}+\frac{5}{3}\theta + 
\frac{(\rho+p\dot{)}}{\rho+p}=\sigma^{\mu \nu}
\frac{\omega_{\mu}\omega_{\nu}}{\omega^2} \label{omega} \;,
\end{equation}
for a perfect fluid with equation of state $p=p(\rho)$, immediately 
leading to the result that if 
$\omega$ is zero at some given initial time, then it will be always zero.  

If we deal with $\omega$ as a small perturbation, the right-hand side of 
Eq.~(\ref{omega}) can be neglected (being of higher order due to the presence 
of $\sigma^{\mu \nu}$), 
so, we get the usual angular-momentum conservation law 
(e.g. Ref.~\cite{coles})
\begin{equation}
\omega (\rho+p)a^5 = {\rm constant} \;, 
\label{loyt}
\end{equation}
which means that, even if we start with a small non-zero $\omega$, it 
will decay with time at a rate which depends on the equation of state 
$p(\rho)$.

We can show indeed that at first order $\omega^{i}$ is related to the curl 
of the fluid velocity
\begin{equation}
\omega^i_{(1)} = - \frac{1}{2 a^2} \epsilon^{ijk} (\partial_j v_k
+\partial_j \chi_k) \, 
\end{equation}
and it is well known that this vortical component is not generated by standard 
perturbation generating processes -- such as inflation -- so it can be safely
set to zero. 

At second order, though, the quantity $\omega^i$ does 
not coincide with the curl of the velocity $\zeta^i$, nor with the quantity 
$\beta^i$, but it also contains squared first-order 
terms \footnote{Note however that for a pressureless self-gravitating fluid 
these extra terms vanish.}, 
\begin{equation}
\omega^i= - \frac{1}{2 a^2}[  (a \beta^i)+ \epsilon^{ijk} ( 3 v_j 
\partial_k\varphi+ v_j v_k')] \, .    \label{omega2}
\end{equation}
This shows that in order for $\omega^i$ to vanish 
$\beta^{ i}$ has to be generated.  
In other words, the 
momentum-conservation 
equation, whose general form for a perfect fluid at second order in the 
Poisson gauge is 
\begin{eqnarray}
&& (v_i+\chi_i)'+(1-3 w){\cal H} (v_i+\chi_i)+
\frac{w}{1+w}\partial_i\delta  \nonumber \\
&&+  \partial_i\phi+\frac{w}{1+w}(3
\varphi 
-\delta )\partial_i\delta
- (2-3 w) \varphi'v_{i} \nonumber\\
&&  -  w  v^{ ,j}_{j}
v_{i}
 + \frac{1}{2}(v^{2})_{,i}+ \frac{1}{2}(\varphi^2)_{,i}
=0 \, ,
\label{wgeneric}   
\end{eqnarray}
leads to a conservation equation for $\omega_i$
\begin{equation}
\omega_i'+3(1-w){\cal H}\omega_i=0 \;, 
\end{equation}
which is of course equivalent to Eq.~(\ref{loyt}) above.
In particular, it can be checked that the direct solution, 
Eq.~(\ref{betaespl}), of Eq.~(\ref{wgeneric})
(with $w=1/3$) coincides with the expression of $\beta_i$ which is found 
using the fact that $\omega^i$ in Eq.~(\ref{omega2}) is always zero. 
This result is given in Eq.~(\ref{betaintegrato}). 

Finally we can understand why the magnetic field is generated by
looking at  Eq.~(\ref{protoneq}). In fact, from there, we see that the
magnetic field is sensitive to the vorticity of the charged 
matter components like protons.  
In a Universe with only self-gravitating pressureless matter, 
this would be conserved, so no magnetic field would be created.
Note also that in the case of pure matter it is possible to show that the 
conserved vorticity vector is $a^3\omega^{(m) \, i} =- a^2\beta^{(m)\, i}/2$
(see Ref.~\cite{mhm}). 
Then, let us consider what happens to a sub-dominant matter fluid in a 
radiation dominated Universe.
We may consider first, as an illustration, the case of a non-interacting 
matter component. Since we are in the radiation era the 
matter components yield a 
negligible contribution to gravity, and the potentials are driven by radiation.
So the quantity $\beta^{(m) \, i}$ is no longer conserved but has a source 
(which is non-zero as long as $\varphi^\prime$ is not-zero and its gradient is
not parallel with that of $\varphi$). In fact it has 
to satisfy (the curl of) the momentum-conservation equation for a 
pressureless fluid, which is  
\begin{equation}
\frac{(a \beta^{(m)}_{i})'}{a}+{\cal H} (\beta^{(m)}_{i}) 
- 2 \frac{\epsilon_{ijk} \partial^j (\varphi'v^{k\, (m)}) }{a}=0 \, .
\label{matter}   
\end{equation}
$$
$$
Finally, accounting for an interacting fluid (and so $v^{(p)} \simeq 
v^{(\gamma)}$ and $\beta^{(p)} \simeq \beta^{(\gamma)}$),
Eq.~(\ref{matter}) is no longer satisfied. However, 
the left-hand side is compensated by the 
rotational part of a non-zero electromagnetic field, as one can see in 
Eq.~(\ref{curlproton}), that we rewrite (neglecting once again the 
diffusion term and taking $v^{(p)} \simeq v^{(e)}$) 
as
\begin{eqnarray}
&& \frac{(a \beta^{(m)}_{i})'}{a}+{\cal H} (\beta^{(m)}_{i}) 
 - 2 \frac{\epsilon_{ijk}\partial^j(\varphi'v^{k \, (m)})   }{a} 
=\nonumber\\
&& =-\frac{e}{m} \frac{(a^2 B_i)'}{a^2} \, .
\label{matter+em}   
\end{eqnarray}

\end{document}